\documentclass[leqno]{article}
\usepackage{amsmath,amsfonts,amssymb,amsthm}
\usepackage[letterpaper,top=2cm,bottom=2cm,left=3cm,right=3cm, marginparwidth=2cm, 
]{geometry}

\usepackage{graphicx}
\usepackage{dcolumn}
\usepackage{bm}
\usepackage{amsthm}

\usepackage{hyperref}
\hypersetup{
	colorlinks=true,       
	linkcolor=blue,          
	citecolor=blue,        
	filecolor=magenta,      
	urlcolor=blue           
}

\usepackage{amssymb}
\usepackage{epstopdf}
\usepackage{graphicx}
\usepackage{subfigure}
\usepackage{epsfig}
\usepackage{latexsym, amsmath}
\usepackage[english]{babel}
\usepackage{amsmath, amstext, amsfonts}
\usepackage{float}
\usepackage{appendix}

\begin{document}

{\centering\LARGE A novel approach to generate attractors with a high number of scrolls \par}
\bigskip

{\centering\large J. L. Echenaus\'ia-Monroy$^{1,2}$ and G. Huerta-Cu\'ellar$^{0,1,2,3}$ \par \footnotetext[0]{Corresponding Author}}

{\centering\itshape {$^{1}$Dynamical Systems Laboratory, CULagos, Universidad de Guadalajara, Centro Universitario de los Lagos, Enrique D\'iaz de Le\'on 1144, Paseos de la Monta\~na, 47460, Lagos de Moreno, Jalisco, M\'exico. \\ $^{2}$Applied Mathematics Division, Instituto Potosino de Investigaci\'on Cient\'ifica y Tecnol\'ogica, Camino a la Presa San Jos\'e 2055, Col. Lomas 4ta. Secci\'on, 78216, San Luis Potos\'i, S. L. P., M\'exico.\\
$^{3}$Department of Physics and Earth Sciences, St.Mary’s University, SanAntonio, TX78228, USA. jose.echenausia@lagos.udg.mx$^1$, g.huerta@lagos.udg.mx$^0$\par}}

\begin{abstract}
	
In this paper, it is presented a novel method for increasing the number of scrolls in a hybrid nonlinear switching system. Using the definition of the \textquotedblleft Round to the Nearest Integer Function\textquotedblright, as a generalization of a PWL function, which is capable of generating up to a thousand of scrolls. An equation that characterizes the grown in the number of scrolls is calculated, which fits to the behavior of the system measured by means of the coefficient of determination, denoted $R^{2}$, and pronounced \textquotedblleft R squared\textquotedblright. The proposed equation is based on obtaining as many scrolls as desired, based on the control parameters of the linear operator of the system. The work here presented provides a new approach for the generation and control of a high number of scrolls in a hybrid system. The results are verified for all the scenarios that the equations covers.

\bf{Multiscroll attractors; High number of scrolls; Round Function; Unstable Dissipative Systems; Coefficient of Determination.}
\end{abstract}

\section{Introduction}

The switched nonlinear systems are mostly associated to the generation of chaotic behaviors with the presence of multiple scrolls in their phase spaces \cite{Eric2010, campos2012attractors, campos2016chaotic}. The study of this kind of systems has presented a great interest in the last three decades of scientific development, due to the endless number of possible applications that these systems can have in different areas of science. The term multiscroll it is referred to the generation of an attractor that has, at least, three scrolls in its state space, unlike the dynamical systems of Lorenz \cite{lorenz1963deterministic} or Chua \cite{chua1992genesis}, which only have attractors of double-scroll. Suykens and Vanderwall \cite{suykens1991quasilinear}, proposed this type of systems at the beginning of the 90's, motivated by the idea of obtaining a richer dynamical system, referring to the number of scrolls, than the system of Chua, on which they were based.

There are several approaches for obtaining systems with multiscrolls, which can be classified by considering their constructions as follows: i) adding break points to the function of Chua \cite{yalcin2000experimental, suykens1997family}, ii) using hysteresis functions \cite{newcomb1986chaos, lu2004generating}, iii) implementing functions of sinusoidal type and, iv) applying Piece-Wise Linear functions (PWL), \cite{yalcin2001n, tang2001generation, chua1986double, suykens1993generation, echenausia2018family}. By means of applying a PWL, an hybrid system is generated, which is characterized by the coexistence of continuos dynamics, such as the state variable of the numercial model, and logical decision making \cite{goebel2009hybrid, ontanon2016analog}. The biggest handicap lies when the expected result entails obtaining a higher number of scrolls in the system.

The generation of simple systems with a high number of scrolls is still an open problem, that's why it is proposed an hybrid system based in the use of Unstable Dissipative Systems (UDS) \cite{Eric2010, campos2012attractors, campos2016chaotic} with the implementation of the Round to Nearest Integer Function as commutation law \cite{huerta2014approach, gilardi2017multistability}, for the generation of attractors with a large number of scrolls, which can be controlled based on a growth equation dependent on the control parameters of the system, thus obtaining a system, numerically, easy to implement.

This work is distributed as follows: the first section of the article contains a brief introduction to the problem where the scientific background is described. Throughout the second section, the basic concepts for designing UDS systems, as well as the mechanism for generating attractors with multiscrolls using PWL functions are described. The mathematical model used, as well as the description of the methodology implemented, are addressed in section number three. The results of the characterization of the model and the construction of the equation that governs the growth in the number of scrolls, are described in the fourth caption. The conclusions about the work are shown at the end of this paper.

\section{Theory} 

\subsection{Unstable Dissipative Systems (UDS)}

It is well known that the generation of attractors with multiple scrolls depends both, on the stability of the generated equilibrium points, as well as on the type of the implemented switching function. It is possible to analyze the stability of the equilibrium points through the theory of Unstable Dissipative Systems (UDS), which describes a variety of three-dimensional systems showing dissipative and conservative components. The coexistence of both components causes the appearance of the so-called attractors with multiscrolls. 

As in previous works \cite{campos2016chaotic, echenausia2018family, gilardi2017multistability}, consider a system of three coupled autonomous differential equations,

\begin{equation}\label{ec. 1}
\begin{array}{c}
\dot{X}=\textbf{AX+Bf(x)},
\end{array}
\end{equation}

\noindent where $\textbf{X}=[x,y,z]^{T}\in R^3$, is the state vector, $\textbf{B}=[b_1,b_2,b_3 ]^T\in R^3$, is a constant position vector, $\textbf{A}=[a_{i,j} ]\in R^{3\times3}$, is a constant matrix, and $\textbf{f(x)}$ is a nonlinear function. The kind of behavior exhibited by the system, is defined by the eigenvalues of matrix $\textbf{A}$, which can generate a great variety of combinations and, therefore, the same diversity of behaviors. 

Considering only the cases in which the system described by \textit{ec. (\ref{ec. 1})} has saddle-node equilibrium points, since they have both, stable and unstable varieties, it is possible to characterize the model in the following way: i) a system it is considered as UDS I, if their equilibrium points are hyperbolic-saddle-node, i.e., one eigenvalue is a negative real one, and the other two are complex conjugated with positive real part, where the sum of the components must be less than zero. This last condition fulfills the dissipative conditions of the system \cite{campos2012attractors, ott1981strange}. ii) By the other side, a UDS II system it is defined as those which has a real positive eigenvalue, and two complex conjugated with negative real part, where the sum of its components it is also negative. 

\subsection{Multiscroll Attractors}

\begin{figure}
	\centering
	\subfigure[]{\includegraphics[scale=0.3]{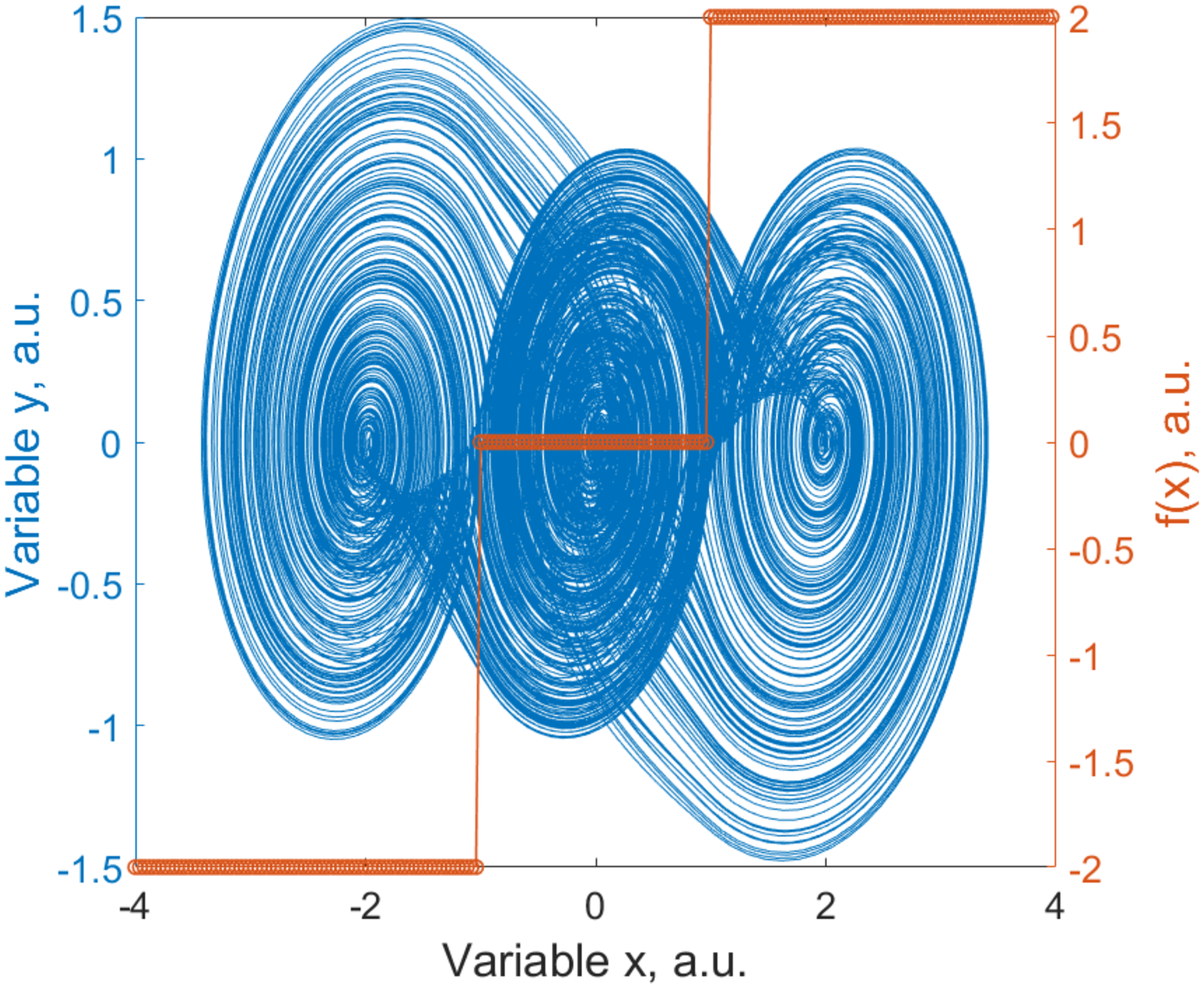}}\hspace{0mm}
	\subfigure[]{\includegraphics[scale=0.3]{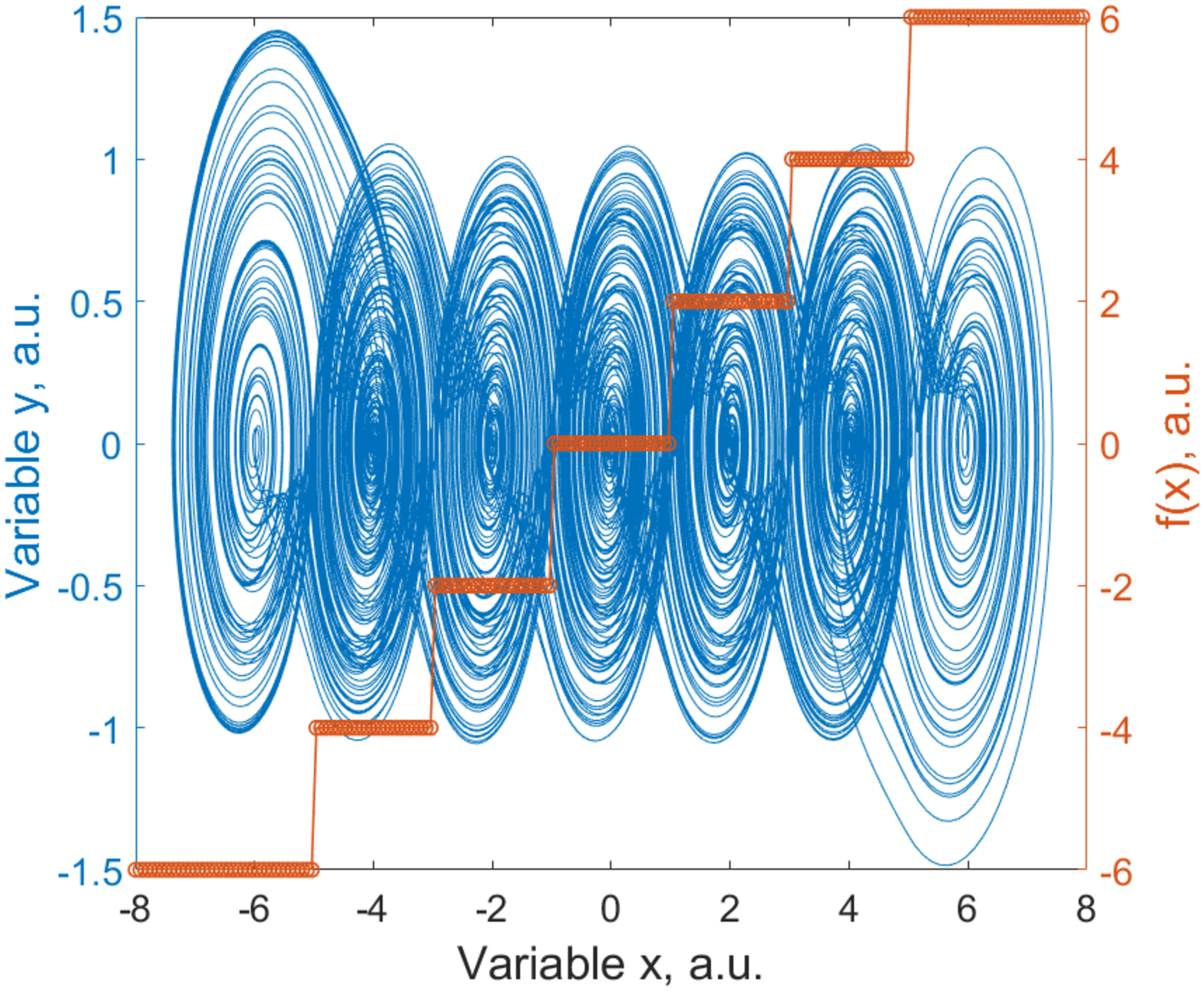}}\hspace{0mm}
	\subfigure[]{\includegraphics[scale=0.3]{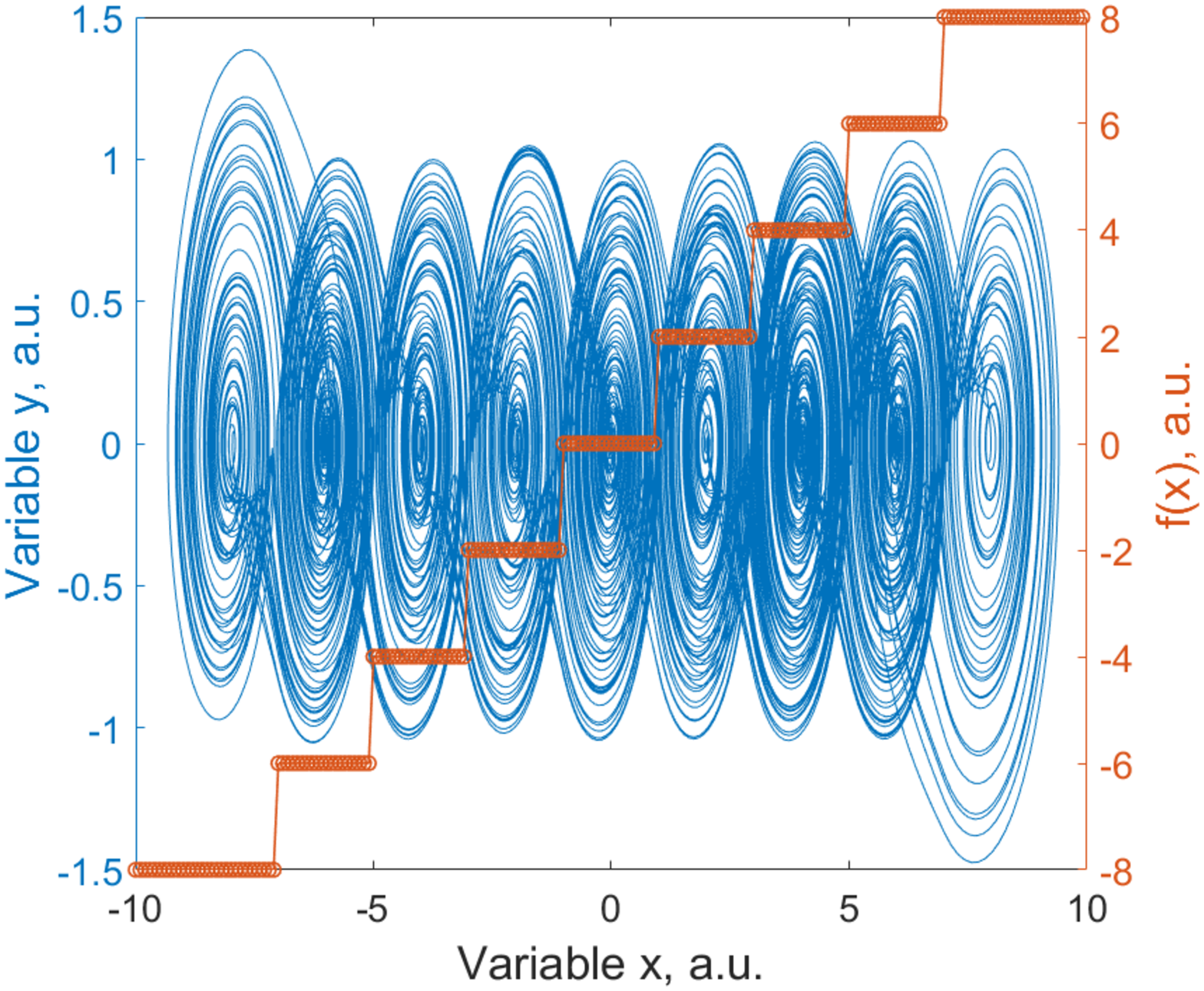}}\hspace{0mm}
	\subfigure[]{\includegraphics[scale=0.3]{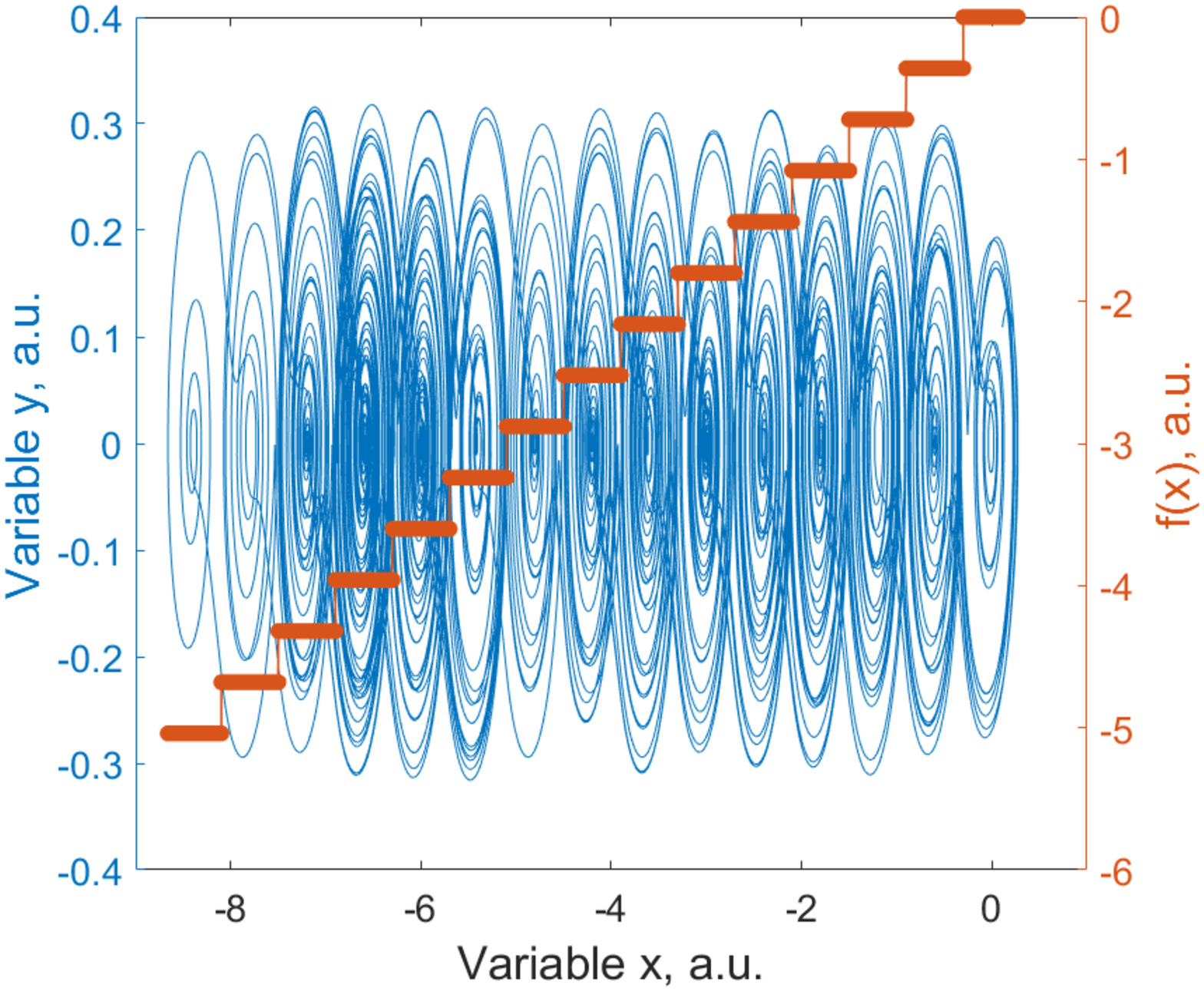}}\hspace{0mm}
	\caption{Attractors generated with the equations (a) \textit{ec. (\ref{ec. 2},3)}, (b) \textit{ec. (\ref{ec. 2},4)}, (c) \textit{ec. (\ref{ec. 2},5)}, (d) \textit{ec. (\ref{ec. 2},6)}.}
	\label{attractor_PWL}
\end{figure}

If the linear operator of the system defined in \textit{ec. (\ref{ec. 1})}, fullfil all the conditions to be defined as a UDS I, then it is possible to generate an attractor with multiple scrolls by means of the construction of a commutation law, in this case, a PWL function. The purpose of the commutation law is control the visit in the different equilibrium points of the system, being achieved by means of the coexistence of a large number of one-spiral unstable trajectories. To illustrate such behavior \cite{Eric2010, campos2016chaotic, echenausia2018family}, consider the following linear operator:

\begin{equation}\label{ec. 2}
\begin{array}{c}
A=\left[\begin{array}{ccc}
\;\;0& \;\;1& \;\;0\\ 
\;\;0& \;\;0& \;\;1\\
-0.6& -0.6& -0.6\end{array}\right],
\end{array}
\end{equation}

\noindent which satisfies the conditions that define a UDS I system, having eigenvalues equal to $\lambda= [- 0.8304,0.3652 \pm 1.2935\;i]$ and $\sum \lambda= -0.1$; this linear operator can be associated to a PWL function of three levels, \textit{ec. (3)}. As can be seen in Figure \ref{attractor_PWL}(a), this conception of the function $f(x)$ generates an attractor of three scrolls with equally distributed trajectories and equidistant equilibrium points. 

If  the same linear operator as the one shown in \textit{ec. (\ref{ec. 2})} is considered, and it is desired to obtain an attractor with a higher number of scrolls, a new commutation law must be built in, taking special care in obtaining equidistant equilibrium points, as well as the distribution of the trajectories. In order to obtain as many scrolls as segments have been introduced to the switching function, i.e., be $f(x)$ the switching function presented in \textit{ec. (4)}, an attractor with seven scrolls in its phase space is obtaining, Figure \ref{attractor_PWL}(b). This same exercise can be done to build in larger switching laws \textit{ec. (5)}, and therefore, generate attractors with the same number of scrolls, Figure \ref{attractor_PWL}(c).

In general, the process to increase the number of scrolls in a system, is exemplified in the set of equations \textit{ec. (3-5)} in Table \ref{tab_pws}, the disadvantage lies when an attractor with many more scrolls must be implemented, i.e. more than a hundred scrolls, which obviously implies a not so simple task to address \cite{tlelo2016generating}.

The task of adding a higher number of equilibrium points to a system, in a simpler way, it is possible through the implementation of a switching function as in \cite{huerta2014approach, gilardi2017multistability}, where the results have been validated for the generation of attractors with multiscrolls. As an example, the Round to the Nearest Integer Function (RNIF) is implemented, \textit{ec. (6)}, which results in the attractor shown in Figure \ref{attractor_PWL}(d).

\begin{table}
	\centering
	\begin{tabular}{c c c}
		\hline \hline
		$\textbf{   \; No. of scrolls} $ & $\textbf{PWL} $ & \\ \hline
		
		\\
		$\textbf{3} $ & 	
		$f(x)= \left\{
		\begin{array}{c}
		-2, \; \mathrm{if}\; x\geq -1;\\
		\;\;\;\;0, \; \mathrm{if}\; -1 <x<1;\\
		2, \; \mathrm{if}\; x\leq 1.
		\end{array} \right.$   & (3)
		\\ \\ \hline \\
		$\textbf{7} $ & 
		$f(x)= \left\{
		\begin{array}{c}
		-6,\; \mathrm{if}\; x\geq -5;\\
		-4, \;\mathrm{if}\; -5 <x<-3;\\
		-2, \;\mathrm{if}\; -3 <x<-1;\\
		0, \;\mathrm{if}\; -1 <x<1;\\
		2, \;\mathrm{if}\; 1 <x<3;\\
		4, \;\mathrm{if}\; 3 <x<5;\\
		6,\; \mathrm{if}\; x\leq 5. 
		\end{array}\right.$ & (4)
		\\ \\ \hline \\
		$\textbf{9} $ & 
		$f(x)= \left\{
		\begin{array}{c}
		-8,\; \mathrm{if}\; x\geq -7;\\
		-6, \; \mathrm{if}\; -7 <x<-5;\\
		-4, \;\mathrm{if}\; -5 <x<-3;\\
		-2, \; \mathrm{if}\; -3 <x<-1;\\
		0, \; \mathrm{if}\; -1 <x<1;\\
		2, \; \mathrm{if}\; 1 <x<3;\\
		4, \; \mathrm{if}\; 3 <x<5;\\
		6, \; \mathrm{if}\; 5 <x<7;\\
		8,\; \mathrm{if}\; x\leq 7. 
		\end{array} \right.$ & (5)
		\\ \\ \hline \\
		
		$\textbf{N} $ & 
		$f(x)= \left\{
		\begin{array}{c}
		\\
		0.36 \left(round\left[ \dfrac{x}{0.6}\right]\right).\\ \\
		\end{array}  \right.$ & (6)\\
		\\ \hline \hline
		\label{tab_pws}	
	\end{tabular}
	\caption{PWL functions used for generate the attractors shown in Figure \ref{attractor_PWL}.}
\end{table}

\subsection{Multiscroll Generator System}

The multiscroll generator system here studied, is similar to those studied in \cite{campos2016chaotic, gilardi2017multistability, echenausiaevidence}, which is composed of three coupled differential equations, and implements a generalization of a PWL function as an approach to obtain multiscroll attractors:

\begin{equation}
\begin{array}{c}
\setcounter{equation}{7}
\dot{x}=y, \\
\dot{y}= z, \\
\dot{z}=-\alpha_{1} x - \alpha_{2} y - \alpha_{3} z + \alpha_{4},\\
\alpha_{4}=C_{1}\left[g\left( \dfrac{x}{C_{2}}\right) \right],
\end{array}\label{modelo}
\end{equation}

\noindent whre the descriptor system to analyze is composed of three state variables $x, y, z$. The values $\alpha_{1,2,3}$ are control parameters of the system that can modify its dynamics, and $\alpha_{4}$ is the switching function associated with the system.  This work is focused on the operation region where the system responds to the configuration of eigenvalues defined as UDS I; so the system is studied under the premise that all control parameters are equal, $\alpha=\alpha_{1}=\alpha_{2}=\alpha_{3}$.

Contemplating the previous statement, the behavior of the characteristic polynomial of the system, $\lambda^3 + \alpha(\lambda^2 + \lambda + 1) = 0 $, is analyzed under the modification for which the model responds to the definition of a system cataloged as UDS I, resulting in a region of operation defined by $ 0 <\alpha < 1 $.

In $\alpha_ {4}$, $C_1$ and $C_2$ are real constants associated with the control parameters of the system. The function $g (x)$ responds as the RNIF, in this case, of the state variable $x$. Because this definition of the round function can be ambiguous for fractional values, the following consideration is adopted:

\begin{eqnarray}\label{round2}
g=\left\{\begin{array}{c} \mathrm{Up\; round,\; by\; taking}\ \,\lfloor x+0.5\rfloor, \\\mathrm{Down\; round,\; by\; taking}\lceil x-0.5 \rceil.\\
\end{array}\right.
\end{eqnarray}

The generalization of the nonlinear function, $\alpha_4$, guarantees the generation of equidistant equilibrium points \cite{gilardi2017multistability}, and the amplitude of the jumps between the different switching surfaces. This conception is similar to those used in the set of equations described by \textit{ec. (3-5)} (Table \ref{tab_pws}), and the result is shown in Figure \ref{attractor_PWL}(d). The nonlinear function implemented in this system has a very similar operation to a PWL function, conceived in order to simplify the construction of many switching surfaces as the control parameters allow.

\section{Methodology and Results}

As it has been demonstrated in previous works \cite{huerta2014approach, gilardi2017multistability, echenausiaevidence}, the use of the nonlinear function $\alpha_{4}$, \textit{ec. (\ref{round2})}, presents a dependence on the control parameters of the system, and on the integration time. This is due that the function does not presents any type of limitation, either temporary or in the number of switching surfaces to visit, so it is defined from $-\infty$ to $\infty$. It is due to this, and the imperative need to generate systems with a higher number of scrolls, that this work is focused on the characterization of the system proposed by \textit{ec. (7)}, to obtain a system of easy numerical implementation, which is able to generate as many scrolls as desired.

According to the considerations raised by the theory that describes the unstable dissipative systems, it was established that the region where the analysis of the proposed model will be performed is defined for values $0<\alpha<1$. Considering future electronic implementations, this control parameter will be explored $0.05\leq\alpha\leq0.95 $, and with a variation in the increment equal to $\Delta_{\alpha}=0.05$.

The system of equations, \textit{ec. (7)}, will be analyzed for each of the values in the control parameter $\alpha$, where the nonlinear function constants will be maintained as follows: $C_{2}=0.6, C_{1}=\alpha C_{2}$, which guarantees the generation of equidistant equilibrium points. This system will be analyzed numerically through the implementation of an integrator type RK4, and a time scale $\tau=0.1$. For each of the $\alpha$ values analyzed, the integration time limit will be gradually increased, and the initial conditions of the system will be randomly changed. For each combination in the parameters $\alpha-t$, the number of scrolls in the model is calculated. An equation describing the growth in the number of scrolls will be approximated, generating a control law to generate attractors with a high number of scrolls.

\begin{figure}
	\centering
	\includegraphics[scale=0.5]{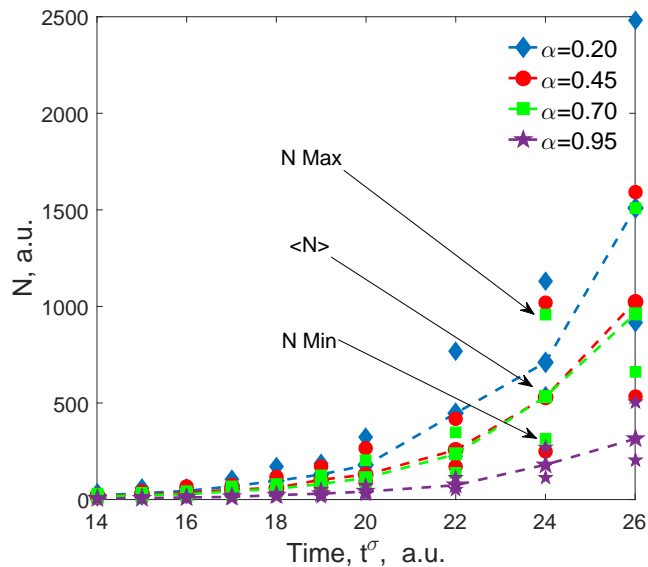}
	\caption{Behavior of the number of scrolls generated, along time, for differents $\alpha$ values. The dotted lines represents the average number of scrolls.}
	\label{Curves}
\end{figure}

In Figure \ref{Curves}, the obtained results from four different control parameters, $\alpha$, and a sample of the temporal values explored are shown. For example, consider $\alpha=0.70$ (green squares), the green dotted line indicates the average of scrolls obtained ($<N>$), with the same marking color, both the maximum ($N$ Max) and minimum ($N$ Min) values are shown. The results are analogous for all the control parameters shown in the figure; turning the graph into a similar one to a box-plot. The temporal spaces analyzed correspond to a value of $2^{\sigma}$, where $\sigma$ is the $x$ axis of the graph. The behavior of the different curves presents the same growth trend, except for the value $\alpha=0.95$, where the increase in the number of scrolls is slower, compared to the rest of the values.

\begin{figure}
	\centering
	\subfigure[]{\includegraphics[scale=0.3]{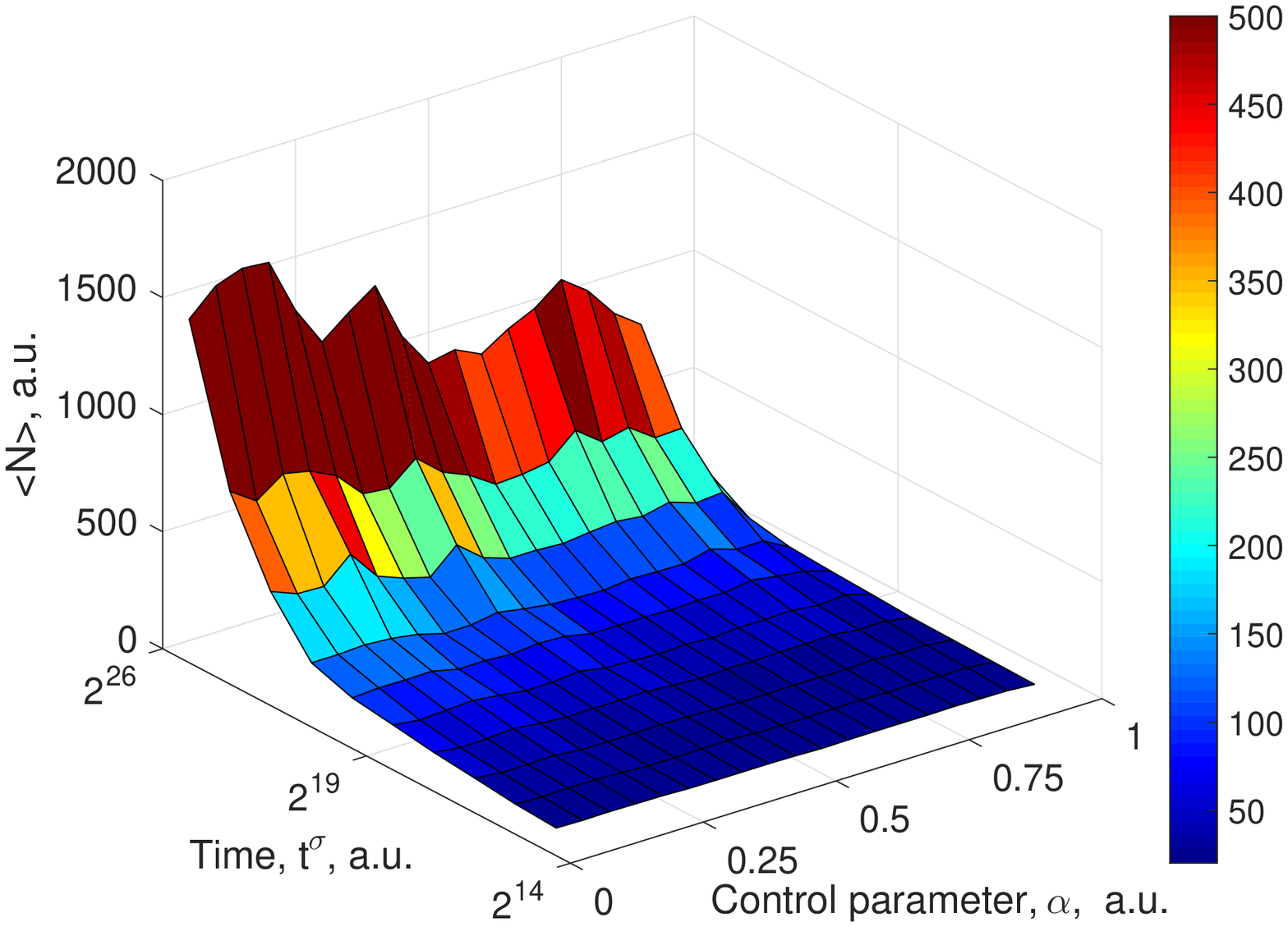}}\hspace{0mm}
	\subfigure[]{\includegraphics[scale=0.3]{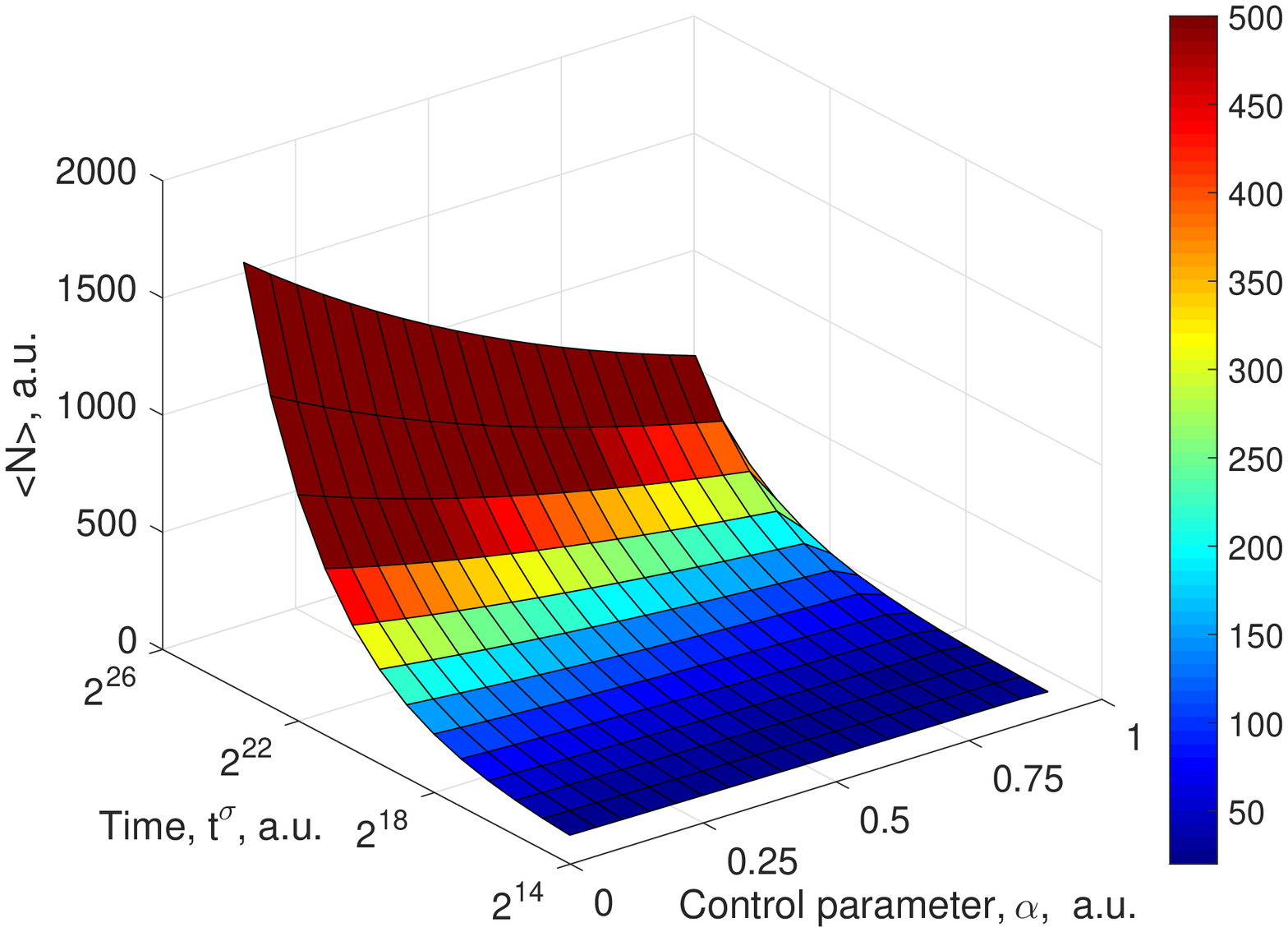}}\hspace{0mm}
	\caption{Surface of behavior in the average number of scrolls generated, along time, for all the $\alpha$ values. (a) Obtained by means of the numerical time series analysis. (b) Obtained by means of applying the \textit{ec. (\ref{ecuacion})}.}
	\label{Surfs}
\end{figure}

Each of the analyzed control parameters yields similar results to those shown in Figure \ref{Curves}, which can be summarized on the surface shown in FIG. 3a), where the increase in the number of average scrolls is shown, $\bar{N}=<N>$, obtained for each of the control parameters with respect to time. Once obtained the results, these can be analyzed to construct an equation that approximates the general behavior of the system. In this case, the adjustment curve is constructed by linearizing the data \cite{mathews2000metodos}, resulting in the equation shown in \textit{ec. (\ref{ecuacion})}, such equation presents a dependency at two parameters, the simulation time given to the system, $t=2^\sigma$, and the control parameter with which it is analyzed, $\alpha$. The behavior of this equation, under the same scenario as those proposed in the numerical analysis, is shown in Figure \ref{Surfs}(b).

\begin{equation}
\begin{array}{c}
\bar{N}=e^{\left[\beta +\dfrac{\ln(t)}{2}\right]},\\
\end{array}\label{ecuacion}
\end{equation}

\noindent where $\beta$ is defined as in \textit{ec. (\ref{beta})}, $n\Delta_{\alpha}=\alpha$, responds to the value in the control parameter that is analyzed, $1\leq n\leq 19$, and $t$ is the time to be simulated in the system, $C_2=0.6$.

\begin{equation}
\begin{array}{c}
\beta=-\Delta_{\alpha}\left[28.2 + n\left(1+\dfrac{3C_{2}}{n} \right) \right] .
\end{array}\label{beta}
\end{equation}

\section{Discussion}

\begin{figure}
	\centering
	\includegraphics[scale=0.5]{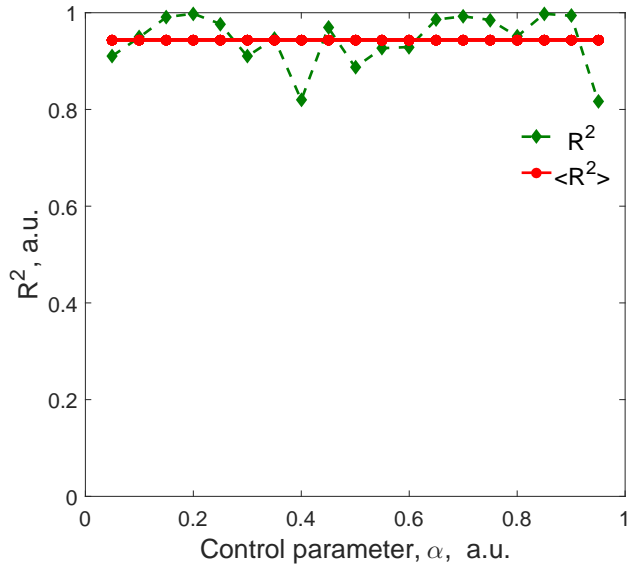}
	\caption{Coefficient of determination ($R^{2}$), between Figure \ref{Surfs}(a) and Figure \ref{Surfs}(b).}
	\label{R2}
\end{figure}

Analyzing the behavior in the growth of the average scrolls, Figure \ref{Surfs}, they seems very similar, but it is not possible to trust in their apparent similarity; that's why the coefficient of determination its calculated for each of the curves experimentally obtained, against the theoretical curve that is generated with the proposed equation. The coefficient of determination, denoted as $R^{2}$, and pronounced \textquotedblleft $R$ squared\textquotedblright, is an statistic tool that determines the quality of a model to replicate results, and the proportion of variation of the proposed results versus the original ones. The accuracy of the prediction depends on the relationship between the variables to be analyzed. The coefficient of determination, \textit{ec. (\ref{r2})}, is defined as the square of the Pearson correlation coefficient; if the result is null, it is said that the predicted variable does not have predictive capacity on the model; as this coefficient increases, the prediction of the model turns out to be more accurate. The $R^2$ values must to be between $0$ and $1$. For each $\alpha$ value, the same time values are analyzed, as in the numerical simulation, once these data are obtained, the coefficient of determination that exists between the two information groups is calculated, resulting in each of the points shown in Figure \ref{R2}.

\begin{equation}
\begin{array}{c}
R^{2}=\left(\dfrac{\sigma_{xy}}{\sigma_{x}\sigma_{y}}\right)^{2},
\end{array}\label{r2}
\end{equation}

\noindent where $\sigma_{xy}$ is the covariance between the original data ($x$) and the proposed regression model ($y$), $\sigma_{x}$ is the standard deviation of the data and $\sigma_{y}$ is the standard deviation of the model.

When its analyzed the behavior of the coefficient of determination, Figure \ref{R2}, it is observed that the equation fits in  a very good way with the experimentally data obtained, where the average value is $\bar{R^{2}}=0.9440$, red line Figure \ref{R2}, confirming the prediction capability of \textit{ec. (\ref{ecuacion})}. The extreme values presented in Figure \ref{R2}, correspond to $\alpha=0.05$ and $\alpha=0.95$. This lack of prediction in the model can be understood from the point of view of the distribution of the recurrence points from the Poincar\'e's section \cite{echenausia2018family}. For very small $\alpha$ values, the disorder in the system is very high, so the visit to the different domains is presented in a highly random way, being more difficult to estimate the number of switching surfaces that will be visit. Similarly for values in the control parameter near to the threshold of the UDS I region, where the system has gained a significant amount of order, so the visit to the different switching surfaces is carried out with a probability that resembles to a Normal distribution, so the prediction in such scenario, also becomes difficult.

\subsection{Equation Validation}

To demostrate the correct operation of the proposed equation, \textit{ec. (\ref{ecuacion})}, the following results are presented, Table \ref{tab_growth}, that cover the three possible scenarios: i) when $\alpha$ and $t$ are known values, determine the number of scrolls to be obtained \textit{ec. (\ref{ecuacion})}, ii) when is desired to obtain an attractor of $N$ scrolls for a determined $\alpha$ value, determine the simulation time that takes the system to reach the visit of the $N$ switching surfaces \textit{ec. (13)}, and iii) when both values, $N$ and $t$, are set parameters, determine the $\alpha$ value that is required to obtain the desired attractor \textit{ec. (14)}.

Considering the first case proposed by \textit{ec. (\ref{ecuacion})}, suppose that the control parameter is known, as well as the simulation time to be used, $\alpha=0.5,\;t=70000 \mathrm{\;u.}$, and it is desired to know the number of scrolls to be obtained, this to determine if the attractor is large enough for the purposes, or a higher number of scrolls is required. With this information, it is possible to estimate the number of scrolls that would be obtained by substituting in \textit{ec. (\ref{ecuacion})}, resulting an average number of scrolls $\bar{N}=35.8064$.

\begin{figure}
	\centering
	\subfigure[]{\includegraphics[scale=0.4]{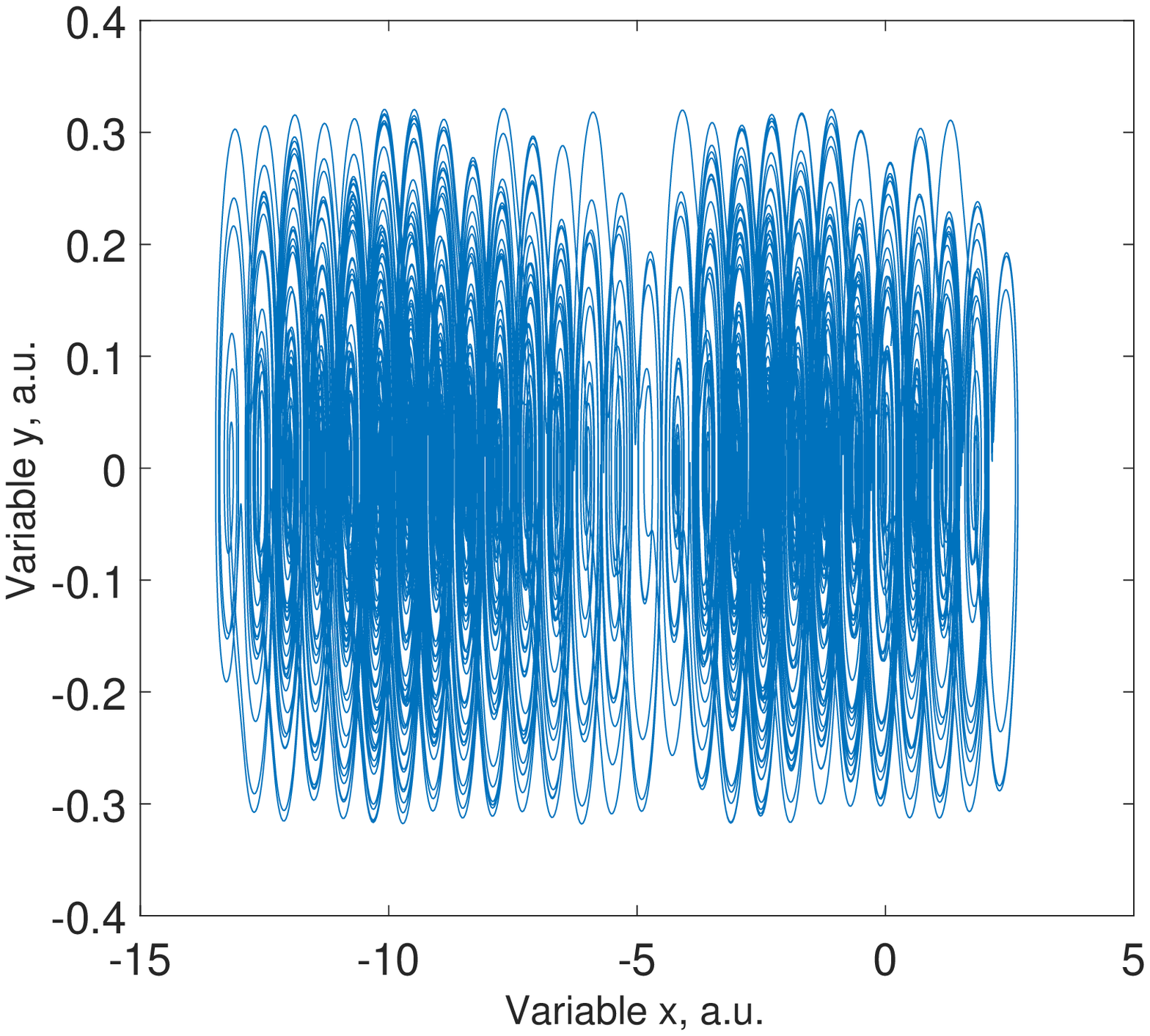}}\hspace{0mm}
	\subfigure[]{\includegraphics[scale=0.4]{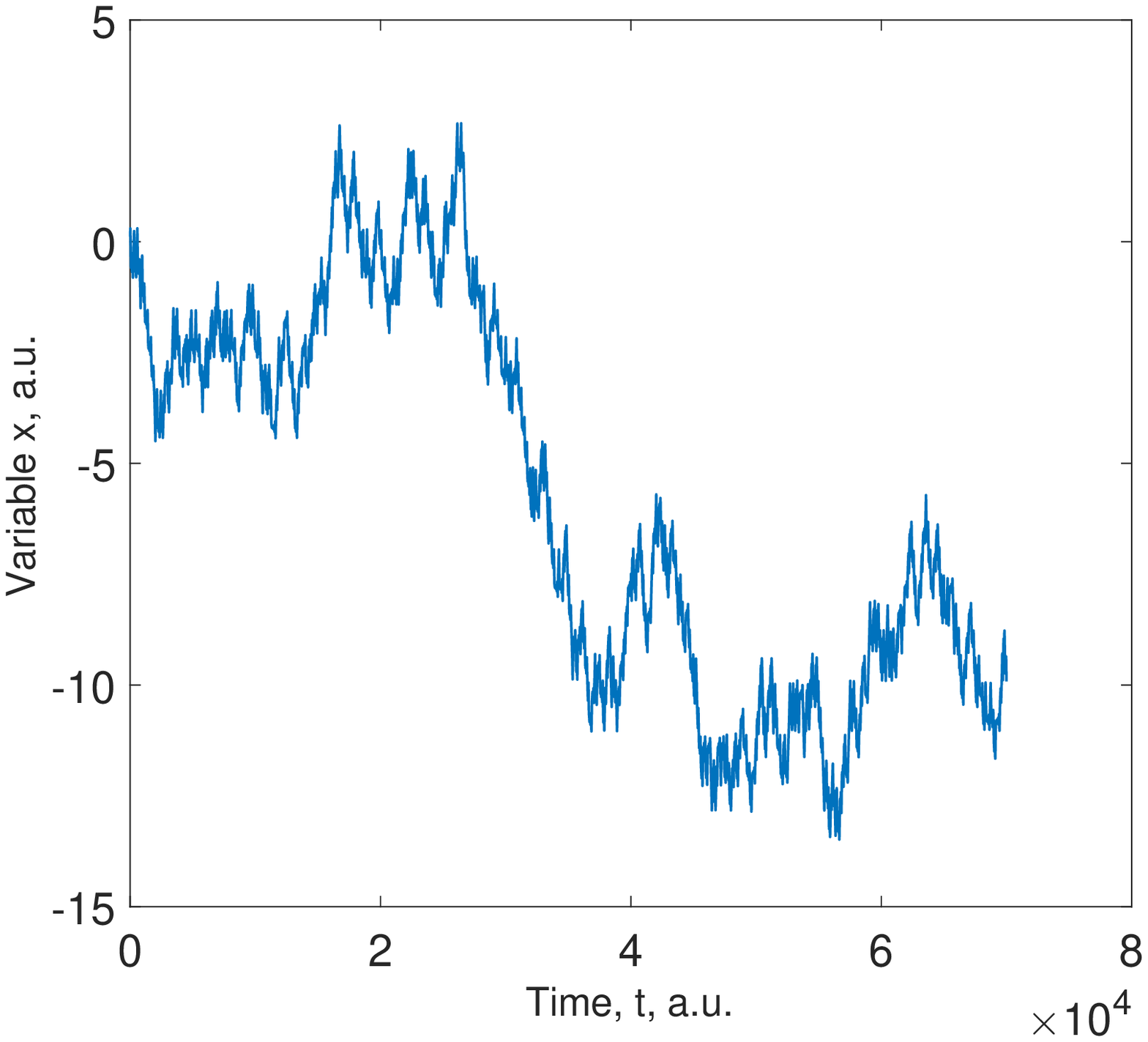}}\hspace{0mm}
	\caption{a) Attractor with $N=27$ scrolls, and b) Time serie of the $x$ variable, for the parameters $\alpha=0.5$, and $t=70000\mathrm{\;units}$.}
	\label{Case_1}
\end{figure}

According to the results of the equation, the system must present around of $36$ scrolls in its phase space, for the parameters $\alpha=0.5$ and $t=70000\mathrm{\;units}$. If this scenario is simulated, the system presents the dynamics shown in Figure. \ref{Case_1}, where an attractor with $27$ scrolls is obtained, which is not so different from the value dessired.

\begin{table}
	\centering
	\begin{tabular}{c c c}
		\hline \hline 
		$\textbf{   \; Known Parameters} $ & $\textbf{Estimated Parameter} $ & \\ \hline
		
		\\
		$\alpha, t$ & 	$
		\bar{N}=e^{\left[\beta +\dfrac{\ln(t)}{2}\right]} $. & (9)\\
		\\	\hline
		\\
		$\alpha, N$ & 	$
		\bar{t}=e^{\left[2\ln\left(\dfrac{N}{e^{\beta}}\right)\right]} $. & (13) \\
		\\ \hline 
		\\
		$t, N$ & 
		$\bar{n}=-(3C_{2}+28.2) -\left[ \ln\left(\dfrac{N}{e^{\left[\dfrac{\ln(t)}{2}\right]}}\right) \left(\dfrac{1}{\Delta_{\alpha}}\right)\right] $. & (14) \\ \\
		\hline \hline 
		\label{tab_growth}	
	\end{tabular}
	\caption{Combinations of the equation for control the growth of the number of multiscrolls.}
\end{table}

\begin{figure}
	\centering
	\subfigure[]{\includegraphics[scale=0.4]{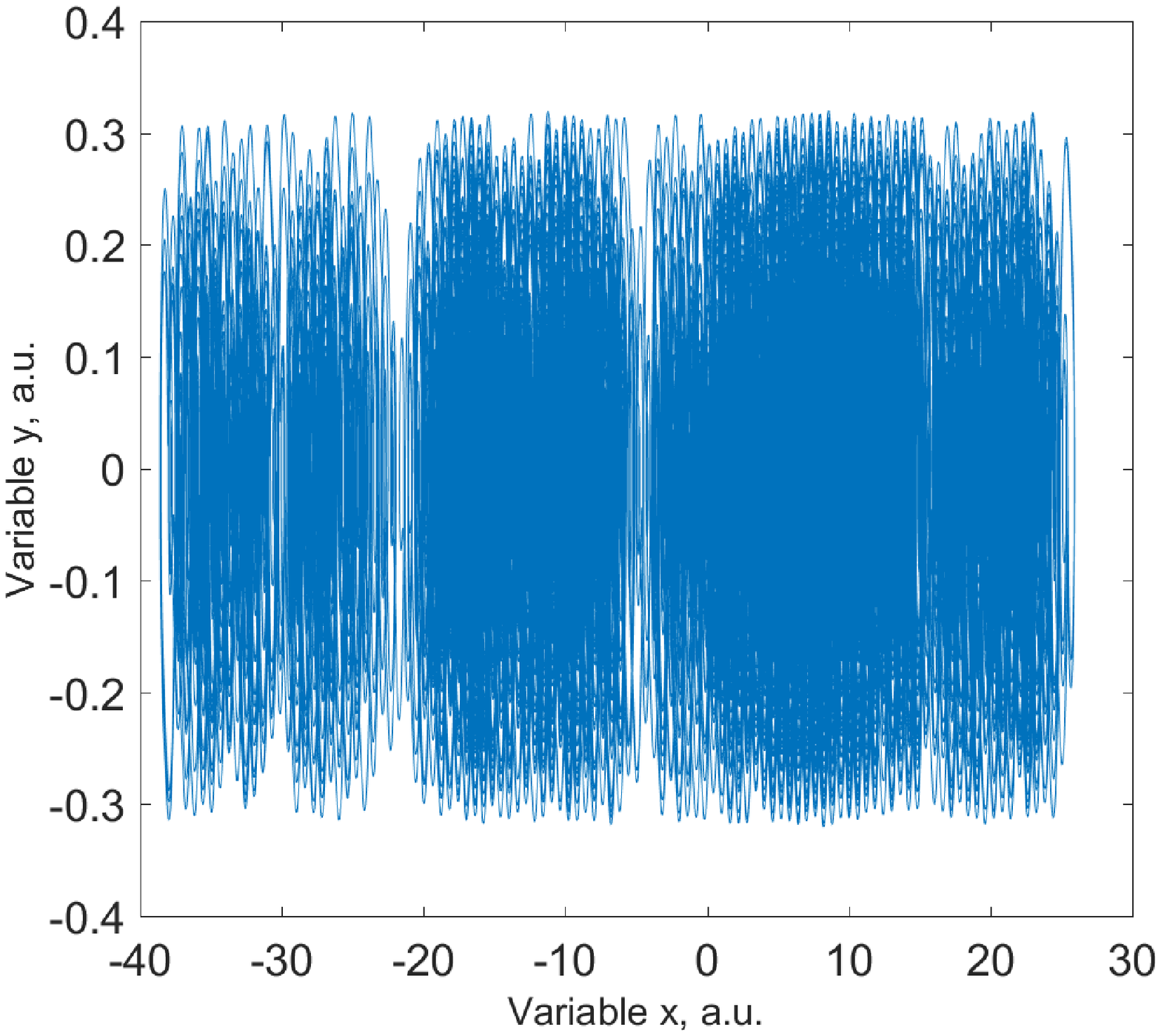}}\hspace{0mm}
	\subfigure[]{\includegraphics[scale=0.4]{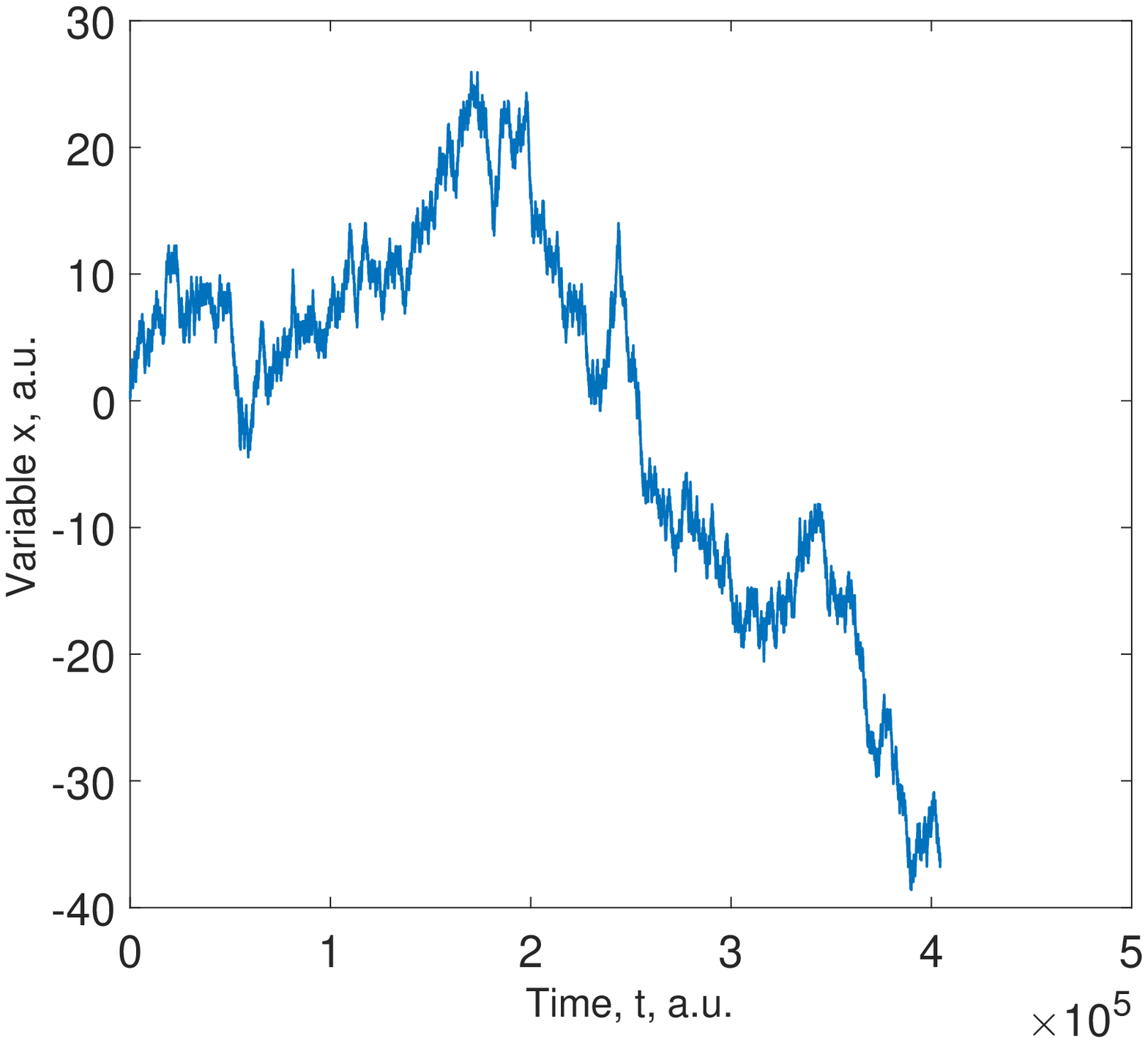}}\hspace{0mm}
	\caption{a) Attractor with $N=108$ scrolls, and b) Time serie of the $x$ variable, for the parameters $\alpha=0.35$, and $t=404470\mathrm{\;units}$.}
	\label{Case_2}
\end{figure}

In the same way as in the previous case, it is possible to perform the following exercise: an attractor with a hundred of scrolls is desired, associated to a control parameter $\alpha=0.35$, so the simulation time needed to obtain such attractor is the parameter to calculate. Substituting these values in \textit{ec. (13)},  $\bar{t}=404470\mathrm{\;u.}$ is obtained. This amount of time may be considered as the minimum time that the system needs to visit all the switching surfaces desired. If the system is simulated with these parameters, an attractor with $108$ scrolls is obtained, Figure. \ref{Case_2}.

\begin{figure}
	\centering
	\subfigure[]{\includegraphics[scale=0.4]{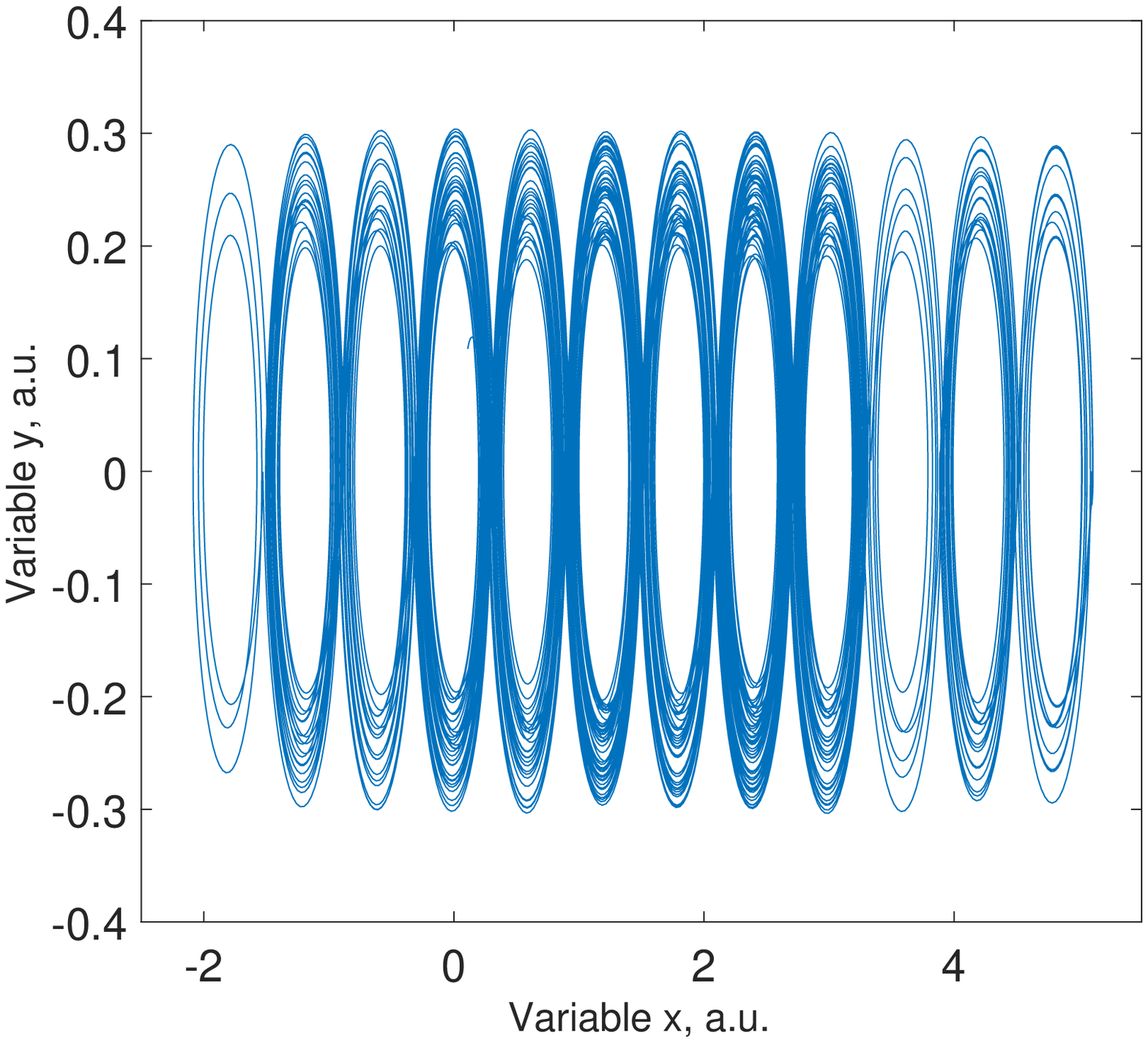}}\hspace{0mm}
	\subfigure[]{\includegraphics[scale=0.4]{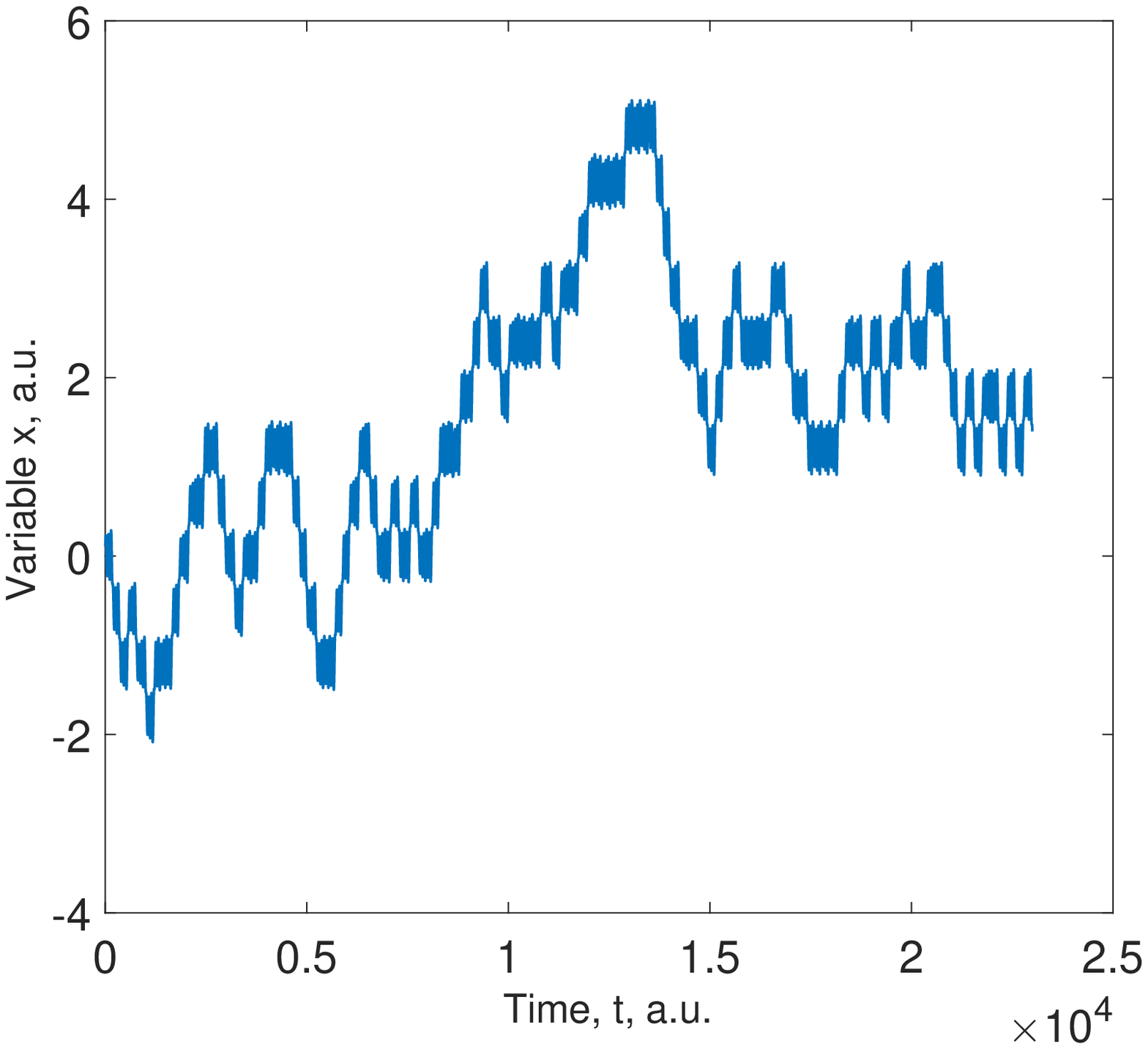}}\hspace{0mm}
	\caption{a) Attractor with $N=12$ scrolls, and b) Time serie of the $x$ variable, for the parameters $\alpha=0.9$, and $t=23000\mathrm{\;units}$.}
	\label{Case_3}
\end{figure}

As a last proof of the right prediction performed by the equation that governs the number of scrolls to be obtained, assume a simulation time $t=23000\mathrm{\;u.}$, and the desired number of scrolls is $N=13$, so the control parameter to be used is unknown. Substituting the information in \textit{ec. (14)}, $\bar{n}=19.1335$ is calculated, which is rounded to $\bar{n}=19$, and corresponds to  $\alpha=0.95$. Carring out the simultaion, the system presents an attractor with only $5$ scrolls. As mention before, the system is hard to predict in this $\alpha$ value, because of the order that the dynamics in the system has gain. Considering this fact, the simulation it is performed for the $\alpha$ previous value ($\alpha=0.9$), where the system exhibits an attractor with $12$ scrolls, Figure \ref{Case_3}.

\section{Conclusions}

With this research work, a hybrid system of three differential equations that implements the Round to the Nearest Integer Function, as a generalization of a PWL function, for obtaining systems with multiscrolls based on UDS models, has been analyzed. Based on the implementation of the RNIF, a characterization of the behavior of the system was carried out, with which an equation is obtained, dependent on the control parameters, that facilitates the implementation of systems with a large number of scrolls. This equation was validated in all the scenarios that it covers, based on the coefficient of determination ($R^{2}$), guaranteeing a good prediction in the increase of the number of scrolls in the system. 

The model here described, furthermore of facilitate the obtention of systems with a high number of scrolls, is capable of generating attractors with a higher level of disorder, based on the results shown in \cite{echenausia2018family}. The limitations in the basins of attraction for UDS I systems are a major factor when designing multiscroll attractors. The implementation of the results here described, may been helpful in the analysis of systems like those studied in \cite{gilardi2017multistability}, developing the analysis of time series with the minimum amount of data to reproduce the phenomenon. Contemplating all these factors, it is considered that the technological applications of the model are even more attractive, having potential use in fields of science like in neural systems, secure communication systems, electric motors with variable torque, pseudo-random number generators, among others. The improvement in the equation and corresponding analogical implementation, it is proposed as future work.

\section*{Acknowledgments}

J.L.E.M. acknowledges CONACYT for the financial support (National Fellowship CVU-706850, No. 582124), to the University of Guadalajara, CULagos (M\'exico), and to E. Campos-Cant\'on for fructifer discussions and the opportunity of realize a research-stay in his investigation group. This work was supported by the University of Guadalajara under the project \textquotedblleft Research Laboratory Equipment for Academic Groups in Optoelectronics from CULAGOS\textquotedblright, R-0138/2016, Agreement RG/019/2016-UdeG, Mexico.


\begin{thebibliography}{100}


\bibitem{Eric2010} Campos-Cant\'on, E., Barajas-Ram\'irez, J. G., Solis-Perales, G., \& Femat, R. (2010). Multiscroll attractors by switching systems. Chaos: An Interdisciplinary Journal of Nonlinear Science, 20(1), 013116.

\bibitem{campos2012attractors} Campos-Cant\'on, E., Femat, R., \& Chen, G. (2012). Attractors generated from switching unstable dissipative systems. Chaos: An Interdisciplinary Journal of Nonlinear Science, 22(3), 033121.

\bibitem{campos2016chaotic} Campos-Cant\'on, E. (2016). Chaotic attractors based on unstable dissipative systems via third-order differential equation. International Journal of Modern Physics C, 27(01), 1650008.

\bibitem{lorenz1963deterministic} Lorenz, E. N. (1963). Deterministic nonperiodic flow. Journal of the atmospheric sciences, 20(2), 130-141.

\bibitem{chua1992genesis} Chua, L. O. (1992). The genesis of Chua's circuit. Berkeley, CA, USA: Electronics Research Laboratory, College of Engineering, University of California.

\bibitem{suykens1991quasilinear} Suykens, J., \& Vandewalle, J. (1991). Quasilinear approach to nonlinear systems and the design of n-double scroll (n= 1, 2, 3, 4,…). IEE Proceedings G (Circuits, Devices and Systems), 138(5), 595-603.

\bibitem{yalcin2000experimental} Yalcin, M. E., Suykens, J. A. K., \& Vandewalle, J. (2000). Experimental confirmation of 3-and 5-scroll attractors from a generalized Chua's circuit. IEEE Transactions on Circuits and Systems I: Fundamental Theory and Applications, 47(3), 425-429.

\bibitem{suykens1997family} Suykens, J. A., \& Huang, A. (1997). A family of n-scroll attractors from a generalized Chua's circuit. Archiv fur Elektronik und Ubertragungstechnik, 51(3), 131-137.

\bibitem{newcomb1986chaos} Newcomb, R. W., \& El-Leithy, N. (1986). Chaos generation using binary hysteresis. Circuits, Systems and Signal Processing, 5(3), 321-341.

\bibitem{lu2004generating} Lu, J., Han, F., Yu, X., \& Chen, G. (2004). Generating 3-D multi-scroll chaotic attractors: A hysteresis series switching method. Automatica, 40(10), 1677-1687.

\bibitem{yalcin2001n} Yalcin, M. E., Ozoguz, S., Suykens, J. A. K., \& Vandewalle, J. (2001). n-scroll chaos generators: A simple circuit model. Electronics Letters, 37(3), 147-148.

\bibitem{tang2001generation} Tang, W. K., Zhong, G. Q., Chen, G., \& Man, K. F. (2001). Generation of n-scroll attractors via sine function. IEEE Transactions on Circuits and Systems I: Fundamental Theory and Applications, 48(11), 1369-1372.

\bibitem{chua1986double} Chua, L., Komuro, M., \& Matsumoto, T. (1986). The double scroll family. IEEE transactions on circuits and systems, 33(11), 1072-1118.

\bibitem{suykens1993generation} Suykens, J. A., \& Vandewalle, J. (1993). Generation of n-double scrolls (n= 1, 2, 3, 4,...). IEEE Transactions on Circuits and Systems I: Fundamental Theory and Applications, 40(11), 861-867.

\bibitem{echenausia2018family} Echenaus\'ia-Monroy, J. L., García-L\'opez, J. H., Jaimes-Re\'ategui, R., L\'opez-Mancilla, D., \& Huerta-Cuellar, G. (2018). Family of bistable attractors contained in an unstable dissipative switching system associated to a SNLF. Complexity, 2018.

\bibitem{goebel2009hybrid} Goebel, R., Sanfelice, R. G., \& Teel, A. R. (2009). Hybrid dynamical systems. IEEE Control Systems Magazine, 29(2), 28-93.

\bibitem{ontanon2016analog} Ontañ\'on-Garc\'ia, L. J., Campos-Cant\'on, E., \& Femat, R. (2016). Analog electronic implementation of a class of hybrid dissipative dynamical system. International Journal of Bifurcation and Chaos, 26(01), 1650018.

\bibitem{huerta2014approach} Huerta-Cuellar, G., Jimenez-Lopez, E., Campos-Cant\'on, E., \& Pisarchik, A. N. (2014). An approach to generate deterministic Brownian motion. Communications in Nonlinear Science and Numerical Simulation, 19(8), 2740-2746.

\bibitem{gilardi2017multistability} Gilardi-Vel\'azquez, H. E., Ontañ\'on-Garc\'ia, L. J., Hurtado-Rodriguez, D. G., \& Campos-Cant\'on, E. (2017). Multistability in piecewise linear systems versus eigenspectra variation and round function. International Journal of Bifurcation and Chaos, 27(09), 1730031.

\bibitem{ott1981strange} Ott, E. (1981). Strange attractors and chaotic motions of dynamical systems. Reviews of Modern Physics, 53(4), 655.

\bibitem{tlelo2016generating} Tlelo-Cuautle, E., Pano-Azucena, A. D., Rangel-Magdaleno, J. J., Carbajal-Gomez, V. H., \& Rodriguez-Gomez, G. (2016). Generating a 50-scroll chaotic attractor at 66 MHz by using FPGAs. Nonlinear Dynamics, 85(4), 2143-2157.

\bibitem{echenausiaevidence} Echenaus\'ia-Monroy, J. L., Garc\'ia-L\'opez, J. H., Jaimes-Re\'ategui, R., \& Huerta-Cuellar, G. Evidence of multistability in a multiscroll generator system.

\bibitem{mathews2000metodos} Mathews, J. H., \& Fink, K. D. (2004). Numerical methods using MATLAB (Vol. 3). Upper Saddle River, NJ: Pearson Prentice Hall.



\end{thebibliography}
\end{document}